# BreakNet: Discontinuity-Resilient Multi-Scale Transformer Segmentation of Retinal Layers

RAZIEH GANJEE[1], BINGJIE WANG[1], LINGYUN WANG[1,2], CHENGCHENG ZHAO[1,2], JOSÉ-ALAIN SAHEL[1], AND SHAOHUA PI[1,2*]

*1 Department of Ophthalmology, University of Pittsburgh, Pittsburgh, PA 15213, USA*
*2 Department of Bioengineering, University of Pittsburgh, Pittsburgh, PA 15213, USA*

*\*shaohua@pitt.edu*

**Abstract:** Visible light optical coherence tomography (vis-OCT) is gaining traction for retinal imaging due to its high resolution and functional capabilities. However, the significant absorption of hemoglobin in the visible light range leads to pronounced shadow artifacts from retinal blood vessels, posing challenges for accurate layer segmentation. In this study, we present BreakNet, a multi-scale Transformer-based segmentation model designed to address boundary discontinuities caused by these shadow artifacts. BreakNet utilizes hierarchical Transformer and convolutional blocks to extract multi-scale global and local feature maps, capturing essential contextual, textural, and edge characteristics. The model incorporates decoder blocks that expand pathwaproys to enhance the extraction of fine details and semantic information, ensuring precise segmentation. Evaluated on rodent retinal images acquired with prototype vis-OCT, BreakNet demonstrated superior performance over state-of-the-art segmentation models, such as TCCT-BP and U-Net, even when faced with limited-quality ground truth data. Our findings indicate that BreakNet has the potential to significantly improve retinal quantification and analysis.

## 1. Introduction

Visible-light optical coherence tomography (vis-OCT) [1-8] is emerging as a powerful tool for retinal imaging due to its ability to produce high-resolution and functional images. By leveraging shorter wavelengths in the visible light spectrum, vis-OCT surpasses standard near-infrared OCT in both detail and contrast. However, vis-OCT faces significant challenges, primarily due to pronounced blood vessel shadow artifacts caused by strong hemoglobin absorption which can provide valuable data for spectroscopic analysis of oxygen saturation but also severely obstruct the OCT reflectance signal beneath the blood vessels, causing discontinuity in other layers and complicating accurate retinal layer segmentation. This issue is particularly problematic in humans, where larger retinal arteries and veins exacerbate the discontinuity, as the degree of signal reduction is proportional to the diameter of the blood vessels.

The first attempt at automated vis-OCT retinal layer segmentation was made by Soetikno et al., using a graph-search technique [9]. Later, they employed a four-level U-Net to segment human retinal images acquired by vis-OCT [10]. Our previous work involved developing an end-to-end deep learning method to segment retinal layers and vascular plexuses using a three-dimensional convolutional neural network (CNN) model for vis-OCT images [11]. While effective, this approach often suffers from errors caused by blood vessel shadows. Recently, Ye et al. introduced a co-learning deep framework using a customized U-Net for simultaneous self-denoising and retinal layer segmentation in vis-OCT images [12]. This model integrates residual connections within the convolutional blocks of each decoder layer to enhance segmentation accuracy. However, the specific challenge posed by blood vessel shadows remains inadequately addressed.

In recent years, vision Transformers [13] have emerged as promising tools for medical image analysis. TransUNet [14], for instance, replaces the CNN encoder in U-Net with a Transformer to encode tokenized image patches, capturing global features. The Swin Transformer employs a pure transformer model with shifted windows during feature upsampling to improve the spatial distribution of feature maps [15]. Leveraging their improved accuracy and efficiency, these models have also been explored for OCT retinal layer segmentation. Tan et al. presented a hybrid model integrating CNN and a lightweight Transformer, processing image inputs through two distinct frameworks: one utilizing the Transformer for global feature extraction and the other using cross-convolution for local feature extraction [16]. They introduced a boundary regression loss function and implemented feature polarization techniques to enhance boundary accuracy and minimize mutual interference during segmentation. Similarly, Cao et al. used A-lines as training data to enhance the multi-head self-attention mechanism of the Transformer [17]. The authors also developed a framework akin to the attention U-Net architecture [18] by combining CNN with Vision Transformer [19].

Building on these recent advances in Transformer-based approaches, we propose BreakNet, a multi-scale convolutional and Transformer-based model specifically designed to address the discontinuity challenges caused by blood vessel shadows in vis-OCT retinal image segmentation. BreakNet employs a hierarchical architecture, leveraging multi-path convolutional and vision Transformer blocks to enhance local and global feature extraction. Our method was validated using vis-OCT rodent retinal images, demonstrating superior segmentation performance and showcasing the potential of vis-OCT in retinal imaging.

## 2. Methods

### 2.1 Vis-OCT retinal image acquisition

Healthy rodents (N=6 brown Norway rats and N=4 C57 mice) retinal images in this study were acquired by a customized vis-OCT prototype [3]. Briefly, the system has a full-width half-maximum bandwidth of 90 nm from 510 to 610 nm, operating at a 50 kHz A-line sampling rate. During the imaging session, rodents were anesthetized with 3% isoflurane and injected with a ketamine and xylazine cocktail (ketamine: 0.37 mg/kg; xylazine: 0.07 mg/kg). A drop of 1% tropicamide hydrochloride ophthalmic solution was used to dilate the pupil. To prevent corneal dehydration, artificial tears were applied to the eyes before every OCT scan acquisition. Each volume contains 500 A-lines per B-scan, 2 repeated frames for each B-scan, and 500 B-scans in total. The interferogram of each scan was recorded by a line scan camera (Basler spl4096-140km) and further processed in MATLAB to resolve the OCT images. Ethics approval for the protocols was obtained from the Institutional Animal Care and Use Committee (IACUC) of the University of Pittsburgh.

### 2.2 Layer Discontinuity Caused by Blood Vessel Shadow

As shown in Fig. 1, vis-OCT retinal images exhibit strong blood vessel shadows, unlike standard near-infrared OCT (NIR-OCT). These shadows block the OCT signal beneath the vessels, causing significant discontinuities in retinal layers. These discontinuities present challenges because retinal layer segmentation algorithms typically assume continuous layers and perform poorly when gaps are present. Unfortunately, these discontinuities occur frequently, especially in B-scan frames near the optic nerve head, and can result in multiple, closely spaced gaps or ultrawide discontinuities when the B-scan is almost parallel to the vascular patterns (Fig. 1). We anticipate that layer discontinuities will pose even greater challenges for segmenting vis-OCT human retinal images due to the larger size of retinal vessels in humans, which leads to a stronger accumulation of signal extinction compared to rodents.

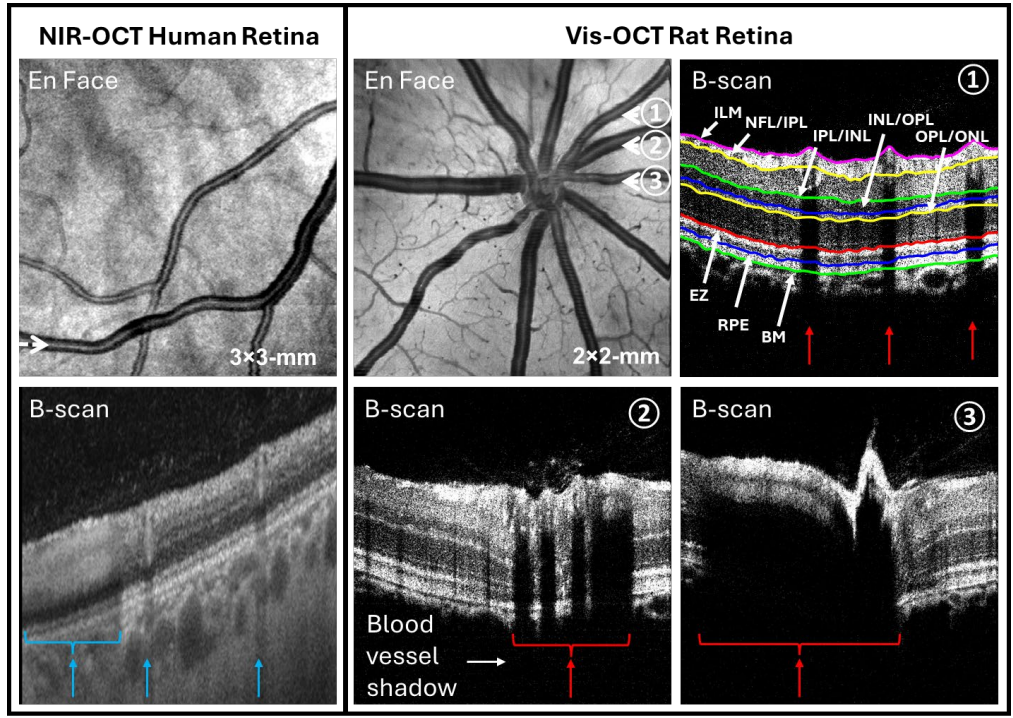

**Fig. 1.** Representative retinal layer discontinuity caused by blood vessel shadows. Left panel: human retinal images acquired with a commercial near-infrared OCT system (Zeiss Plex Elite at λ=1050 nm) showing minimal blood vessel shadows and neglecting retinal layer discontinuity. Right panel: rat retinal images acquired with visible light OCT (vis-OCT) prototypes showing significant blood vessel shadows and layer discontinuity with various situations (B-scan #1: normal discontinuity, B-scan #2: multiple discontinuities close to each other, B-scan #3: ultrawide discontinuity caused by parallel B-scan direction with vascular pattern, as well as the cross-section of optic nerve head (merge points of retinal arteries and veins). White arrows in en face images indicated the position of B-scans. Blue (NIR-OCT) and red (vis-OCT) arrows indicated the position of blood vessel shadows and layer discontinuities. ILM: inner limiting membrane. NFL: nerve fiber layer. IPL: inner plexiform layer. INL: inner nuclear layer. OPL: outer plexiform layer. ONL: outer nuclear layer. EZ: ellipsoid zone. RPE: retinal pigment epithelium. BM: Bruch's membrane.

## *2.3 BreakNet*

Inspired by the work presented in [20], we propose a multi-scale feature extraction block (MSFE) enhanced with convolutional and vision transformer elements that incorporate multiple fields of view. This design aims to achieve precise segmentation performance for visible-OCT retinal images. The MSFE block efficiently combines local and low-level features, such as edges and textures, with global and high-level features, like shapes and sizes, simultaneously. We developed our method by integrating this block into a four-stage hierarchical framework, as illustrated in Fig. 2. The backbone architecture begins with a stem convolution layer that includes two 3×3 convolution operations, which help prevent detail loss due to transformer patches. This is followed by a stack of four MSFE blocks, each generating a comprehensive feature map. Our MSFE block consists of two main components: patch embedding and feature extraction, represented by the pink, blue, and green boxes in Fig. 2. The details of each component are described below:

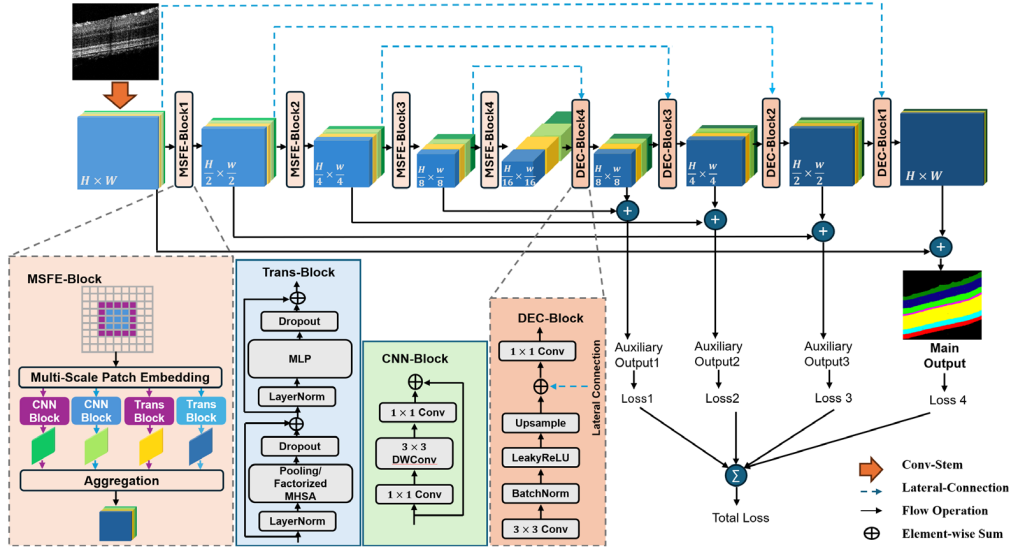

Fig.2. The architecture of the proposed method for deep learning segmentation of retinal layers in visible OCT B-scan images. A B-scan captured from the rat eye is shown as a representative example.

### 2.3.1 Patch embedding

Different patch sizes enable vision transformers to have a multi-scale range of global views, resulting in more efficient semantic feature extraction and consequently more accurate segmentation. While the hierarchical framework of the proposed method already facilitates this capability for transformers, integrating multi-scale global feature extraction at each stage within this hierarchy enhances the power of vision transformers, especially in segmenting challenging images like B-scans with wide discontinuity. To achieve this, we incorporated two scales, 3×3 and 5×5, in patch embedding. This approach enabled us to extract feature maps of the same size using different patch sizes within the same feature layer, creating a multi-scale global feature map at the same level.

In the implementation of the patch embedding layer, the network learned the 2D depth convolutional function $W_{k \times k}^D$ to map the feature map $X_{H_{i-1} \times W_{i-1} \times C_{i-1}}^{i-1}$ from $(i-1)^{th}$ stage into the new token $\tilde{X}_{\frac{H_{i-1}}{2} \times \frac{W_{i-1}}{2} \times C_{i-1}}^i$ in the $i^{th}$ stage using Eq. (1).

$$\tilde{X}_{\frac{H_{i-1}}{2} \times \frac{W_{i-1}}{2} \times C_{i-1}}^i = W_{k \times k}^D(X_{H_{i-1} \times W_{i-1} \times C_{i-1}}^{i-1}), \ k \in \{3,5\} \ and \ i \in \{1,2,3,4\} \quad (1)$$

Here, $k$ is the kernel size or patch size, and $H$, $W$, and $C$ indicate the height, width, and channel sizes of the feature map, respectively. According to this equation, in each stage, two convolutional patch embedding layers with kernel sizes of $3 \times 3$ and $5 \times 5$ were applied to the input feature map to build two same-size feature maps with half of the width and height compared to the previous stage. These feature maps are further proceeded by CNN and transformer blocks to extract local and semantic information, respectively.

### 2.3.2 Feature extraction

In this part of the MSFE block, a multi-scale combination of local and contextual retinal layer features is extracted from the patches by leveraging self-attention mechanisms in transformer blocks (Trans Block) and the inductive bias strength in convolutional layers (CNN Block). Trans blocks enable the network to capture the inter-dependency of features in a long-range global view, and CNN blocks enable the network to extract low-level information like texture and edges.

The MSFE block utilizes two transformers and two CNN blocks, as represented by the blue and green boxes in Fig. 2. CNN blocks 1 and 2 share identical architecture, processing $\tilde{X}^i_{3\times3}$ and $\tilde{X}^i_{5\times5}$ provided by the patch embedding layer via a sequence of $1\times1$ pointwise convolution ($W^P_{1\times1}$) and $3\times3$ depth-wise convolution ($W^D_{3\times3}$), as follows:

$$\dot{X}^i_{k\times k} = W^P_{1\times1}\left(W^D_{3\times3}\left(W^P_{1\times1}(\tilde{X}^i_{k\times k})\right)\right) + \tilde{X}^i_{k\times k}, k \in \{3,5\} \text{ and } i \in \{1,2,3,4\} \quad (2)$$

Trans blocks 1 and 2 also have the same structure comprising of normalization ($Norm$), self-attention ($SA$), dropout ($Dpt$), and multi-layer perceptron ($MLP$) layers. However, two different strategies are employed within their attention layer to reduce the complexity. In Brief, in each stage $i$, the implementation of the Trans blocks for two input feature maps $\tilde{X}^i_{3\times3}$ and $\tilde{X}^i_{5\times5}$ can be summarized as follows:

$$\ddot{X}^i_{k\times k} = Dpt\left(MLP\left(Norm(\bar{X}^i_{k\times k} + \tilde{X}^i_{k\times k})\right)\right) + \bar{X}^i_{k\times k} + \tilde{X}^i_{k\times k}, k \in \{3,5\}, i \in \{1,2,3,4\} \quad (3)$$

$$\bar{X}^i_{k\times k} = Dpt\left(SA\left(Norm(\tilde{X}^i_{k\times k})\right)\right) \quad (4)$$

The self-attention layer in the trans blocks allows the network to achieve a larger receptive field, enabling efficient perception of semantic features. The proposed segmentation model needs these features to learn the varying width and shape of vessel shadow within the layered structure of the retina. This understanding enables accurate segmentation in such B-scans. However, the self-attention layer is a key factor in the high complexity of Transformers. To mitigate this complexity, we replaced the self-attention layer with an average-pooling layer in Trans block 1 [21] for the input feature map $\tilde{X}^i_{3\times3}$; and used a factorized attention layer with linear complexity in Trans block 2 [22] for the input feature map $\tilde{X}^i_{5\times5}$. Accordingly, the SA function in Eq. 4 is described as:

$$SA(Q,k,v) = \begin{cases} AvgPool, & if\ k = 3 \\ \frac{q}{\sqrt{c}}(softmax(k)^T v), & if\ k = 5 \end{cases} \quad (5)$$

Using Eq. 3 and Eq. 4 in each stage $i$, two global feature maps $\ddot{X}^i_{3\times3}$ and $\ddot{X}^i_{5\times5}$ are generated by Trans blocks 1 and 2, respectively, capturing efficient information of long-range dependencies across two different patch sizes. To fuse two local feature maps from CNN blocks, and two contextual feature maps resulting from Trans blocks, a concatenation layer followed by a pointwise convolution is employed as indicated by Eq. 6. :

$$\hat{X}^{i+1} = W^P_{1\times1}\left(concat(\dot{X}^i_{3\times3}, \dot{X}^i_{5\times5}, \ddot{X}^i_{3\times3}, \ddot{X}^i_{5\times5})\right) \quad (6)$$

### 2.3.3 Lateral connection and decoder

To obtain the segmentation probability map, we developed a hierarchical decoder pathway comprising DEC-blocks, as depicted in Fig. 2. In the decoding pathway, higher-resolution feature maps are generated from lower-level features that are spatially coarser but semantically stronger, achieved by incorporating up-sampling layers. However, these multiple down-sampling and up-sampling operations can lead to a loss of precise location information in the feature maps. To address this, lateral connections are employed to enhance the feature maps. Specifically, each up-sampled output in the DEC blocks was combined with the corresponding MSFE block using element-wise addition.

As shown in the architecture of DEC-blocks in Fig. 2, each decoder block comprises five main layers arranged sequentially: a 3×3 depth-wise convolution ($W^D_{3\times3}$), batch normalization ($BN$), leaky ReLU activation function ($LR$), up-sampling layer ($UP$), and $1\times1$ pointwise convolution ($W^P_{1\times1}$). Each DEC Block receives two feature input maps, one from the preceding decoder block output ($D^{i+1}_{H_{i+1}\times W_{i+1}\times C_{i+1}}$) and the other from the corresponding MSFE or conv-steam output ($\hat{X}^i_{H_i\times W_i\times C_i}$) via lateral connections. The interaction between these layers generating decoder output $D^i_{H_i \times W_i \times C_i}$ using two inputs $D^{i+1}_{H_{i+1} \times W_{i+1} \times C_{i+1}}$ and $\hat{X}^i_{H_i \times W_i \times C_i}$ in each stage $i$ can be indicated as follows:

$$D^i_{H_i \times W_i \times C_i} = W^P_{1\times1}((out^i_{H_i \times W_i \times C_i} + \hat{X}^i_{H_i \times W_i \times C_i})) \tag{7}$$

$$out^i_{H_i \times W_i \times C_i} = UP\left(LR\left(BN\left(W^D_{3\times3}(d^{i+1}_{H_{i+1} \times W_{i+1} \times C_{i+1}})\right)\right)\right) \tag{8}$$

Where $H_i = 2 * H_{i+1}$, $W_i = 2 * W_{i+1}$, and $C_i = C_{i+1}$ show the height, width, and channel sizes of feature maps in each stage $i$, respectively.

By iterating the decoding process using four DEC blocks, we computed the final segmentation probability map along with three auxiliary outputs. These auxiliary outputs enable the network to better understand the misrepresentation of feature maps, resulting in a more accurate segmentation. To this end, as illustrated in Fig. 1, each DEC block output $D^i_{H_i \times W_i \times C_i}$ was first summed with its corresponding lateral feature map $\hat{X}^i_{H_i \times W_i \times C_i}$, and then reshaped to match the height and width of the original input image through bilinear interpolation. Subsequently, these three auxiliary outputs, along with the main output derived from the final DEC block, were passed through the SoftMax function to compute the loss function across the ground truth in each layer.

### 2.3.4 Implementation Details

Our rodent dataset consists of 4096 B-scans for training on rats and 2048 B-scans for training on mice. Two trained experts manually provided ground truths for the eight boundaries shown in Fig. 1. Our implementation is based on PyTorch framework and Python 3.8, executed on a PC equipped with an NVIDIA GeForce 4090 and 94 GB of RAM. Data augmentations, including transpose, contrast adjustment, and vertical/horizontal flipping, were applied to enlarge the training sets. Dice loss was used as a loss function. The hyperparameters for training were set as follows: a learning rate of 1e-2 with decay applied every 5 epochs (reducing the learning rate by 0.8), a batch size of 8, a maximum of 60 epochs, and the Adam optimizer.

We conducted our evaluation by considering both computational and clinical perspectives, through qualitative and quantitative analyses. In computational analysis, we utilized three widely recognized measures: the Dice coefficient (Dice), Intersection over Union (IoU), and counter-error (CE) [23, 24]. For clinical analysis, we employed the thickness error (TE) metric. These metrics were computed based on the image label and boundary profile of the intersection of two ground truths. The efficiency of the proposed method was evaluated against, a) <u>T</u>ightly combined <u>C</u>ross-<u>C</u>onvolution and <u>T</u>ransformer with <u>B</u>oundary regression and feature <u>P</u>olarization (TCCT-BP), a leading human retinal segmentation model [16], and b) classical U-Net [25] as a pure CNN model. All methods were trained from scratch on the training set for a fair comparison.

### 3. Results

### 3.1 Overall of BreakNet Layer Segmentation Performance

Owing to the incorporation of MSFE blocks for effective global features extraction, the proposed BreakNet successfully segmented rat retinal images with varying degrees of discontinuity (Fig. 3). In contrast, the U-Net and TCCT-BP models perform well when there is no/minimal discontinuity. Their robustness reduces in images with normal discontinuity caused by retinal major vessels and completely fails in cases with multiple or ultrawide discontinuities. These segmentation errors can happen not only within the shadowed region but also in the neighboring regions.

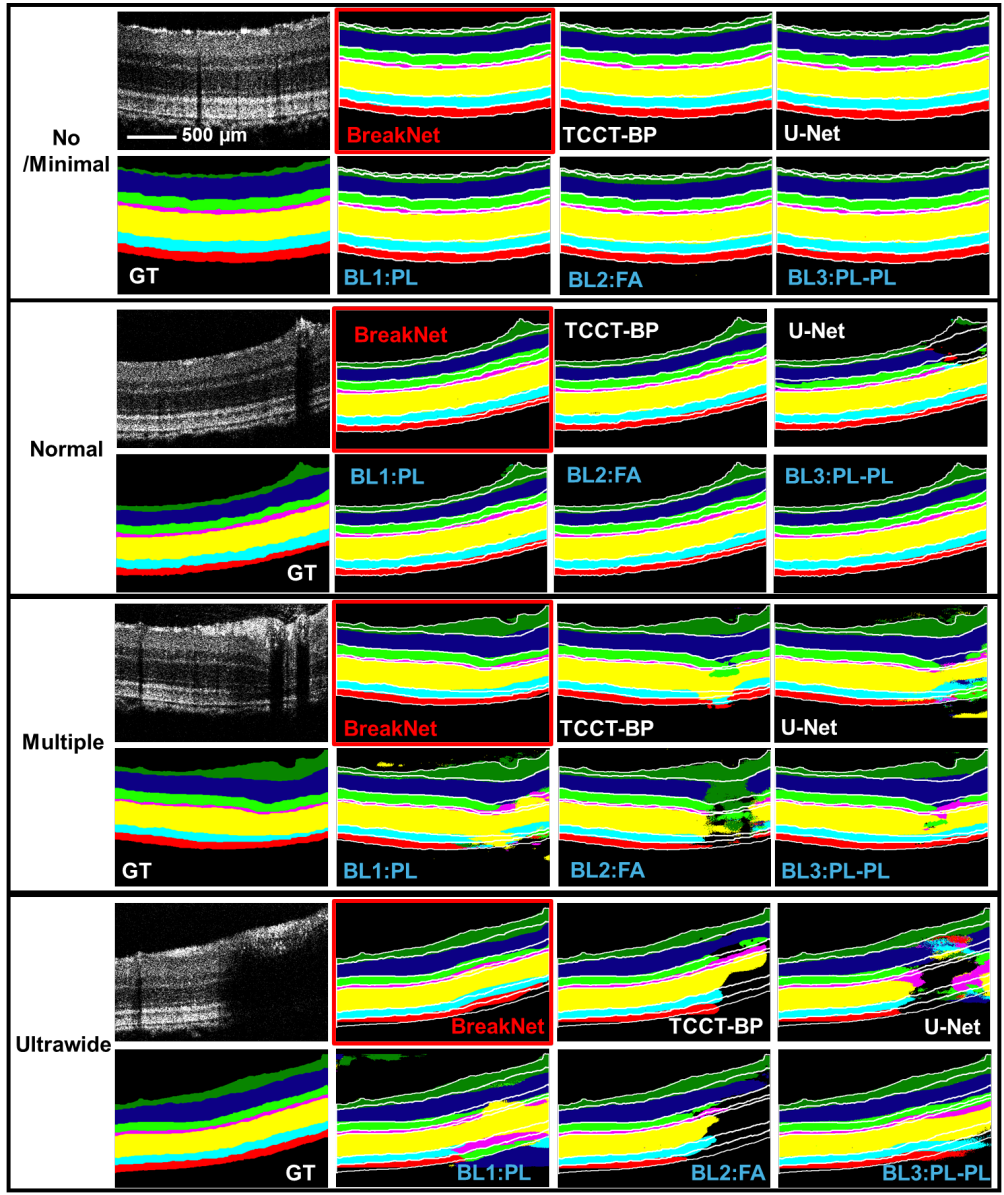

**Fig. 3.** Representative OCT B-scan images of rats, showing the ground truth of 7 retinal layers and subsequent segmentations using the BreakNet, U-Net, TCCT-BP, and baselines. The OCT B-scans are arranged from low to high difficulty in layer segmentation: a regular scan, scans with multiple close shadows, and Ultrawide shadow at the edge. White lines outline the boundary of these layers in the corresponding ground truth.

The Dice and IoU scores for the BreakNet were 0.90 and 0.83 respectively, outperforming those of the U-Net (Dice: 0.81, IoU: 0.72) and TCCT-BP models (Dice: 0.85, IoU:0.76) (Tab. 1). The failure rate of B-Scan segmentation, defined as the proportion of B-scans with segmentation results containing discontinued boundaries compared to the total number of B-scans tested (Tab. 1), was less than 1% (N= 6 / 1141) for BreakNet. In contrast, the failure rate is 38% for U-Net and 11% for TCCT-BP. Excluding the failed B-scans (which are not able to calculate the contour error and thickness error), BreakNet also achieved the best performance on metrics of contour error (2.13 ±0.87 μm) and thickness error (1.61 ±0.73 μm).

Table 1. Quantitative comparison of proposed BreakNet to U-Net and TCCT-BP models on segmenting vis-OCT rat retinal images.

| Method | Dice | IoU | Failure Rate (%) | CE (μm) | TE (μm) |
|---|---|---|---|---|---|
| U-Net | 0.81(0.06) | 0.72(0.08) | 38% | 4.48(1.54) | 4.63(1.81) |
| TCCT-BP | 0.85(0.06) | 0.76(0.08) | 11% | 2.34(1.45) | 2.75(2.31) |
| **BreakNet** | **0.90(0.04)** | **0.83(0.05)** | **1%** | **2.13(0.87)** | **1.61(0.73)** |

## 3.2 Ablation Experiment

To evaluate the necessity of the framework presented MSFE blocks, we modified the architecture of BreakNet by adjusting the number of feature extraction paths (N=1, 2) and attention strategies (Pooling=PL, Factorized Attention: FA) within the Transformers. As shown in Tab. 2, the BL1 and BL2, using single path features, performed poorly in segmenting retinal layers. BL2 performed better than BL1 by replacing PL pooling layers with more complicated and effective FA attention mechanisms. BL3 improved the shape and layer order learning through a dual-path approach for local and global feature extraction but struggled with thick vessel shadows. Our proposed method, which employs a dual-path approach with transformers using both pooling and factorized attention, achieves the best segmentation results (Dice:0.90, IoU: 0.83). It should be mentioned that complicating the segmentation model by employing FA in both paths in BL4 significantly increased the training time (two times than BreakNet) but did not improve performance (Dice:0.90, IoU: 0.83), indicating that the long-distance dependencies captured by FA in the larger patch size already encompass the global information computed in smaller patches by pooling layers.

Table 2. Computational segmentation performance in terms of Dice and IoU (mean (std)) for the proposed method and Baselines on vis-OCT rat retinal images.

| Method | Number of Paths: Patch Size | Attention Strategy | Dice | IoU |
|---|---|---|---|---|
| BL1 | 1: [3× 3] | PL | 0.85(0.06) | 0.77(0.08) |
| BL2 | 1: [3× 3] | FA | 0.87(0.04) | 0.79(0.05) |
| BL3 | 2: [3× 3], [5 × 5] | PL-PL | 0.89(0.05) | 0.81(0.06) |
| BL4 | 2: [3× 3], [5 × 5] | FA-FA | 0.90(0.04) | 0.83(0.05) |
| **BreakNet** | 2: [3× 3], [5 × 5] | PL-FA | **0.90(0.04)** | **0.83(0.05)** |

## 3.4 Robustness to Low-Quality Ground Truth

Preparing high-quality ground truth in medical image segmentation is time-consuming, which can limit the development of deep-learning-based approaches. Therefore, segmentation models that can work with limited-quality ground truth are highly valuable. To evaluate BreakNet's capability in this aspect, we trained the model from scratch using limited-quality ground truth data. Specifically, we manually labeled ground truth on every fifth B-scan and obtained the labels for the remaining B-scans within the volume by interpolation, which is a regular process and causes inevitable displacement of the boundaries, and therefore low quality delineation due to factors such as motion and retinal curvature. As shown in Tab.3, the low-quality ground truth ($GT_L$) generated by interpolation demonstrated a significant difference to the high-quality ground truth ($GT_H$), which was obtained by manually delineating every B-scan. The segmentation results ($SEG_L$) were evaluated using both sets of ground truths data, yielding a

Dice value of 0.83 with $GT_L$, and an improved Dice value of 0.88 with $GT_H$. These values indicate a high correlation of the resulting segmentation with high-quality ground truths, highlighting BreakNet's efficient architecture for learning both local and semantic information, making it robust to noisy ground truth. To further clarify this capability, we inspected the segmentation results in detail (Fig. 4). By comparing the raw B-scans and the segmentations, we found that BreakNet effectively identified the correct position of the layers, despite the evident errors and deviations in the limited-quality ground truth. This performance is close to the 0.90 Dice score obtained when trained with high-quality ground truth, underscoring BreakNet's robustness and accuracy.

Table 3. Evaluation of BreakNet's robustness to training on low-quality ground truth using Dice. $GT_L$: low-quality ground truth; $GT_H$: high-quality ground truth; $SEG_L$: deep learning segmentation using models trained by $GT_L$.

|  | NFL | IPL | INL | OPL | ONL | EZ | RPE | All |
|---|---|---|---|---|---|---|---|---|
| $GT_L$ $GT_H$ | 0.83(0.12) | 0.91(0.05) | 0.85(0.10) | 0.66(0.19) | 0.94(0.04) | 0.86(0.12) | 0.84(0.14) | 0.84(0.10) |
| $SEG_L$ $GT_L$ | 0.82(0.11) | 0.89(0.05) | 0.84(0.10) | 0.66(0.17) | 0.94(0.04) | 0.85(0.11) | 0.83(0.14) | 0.83(0.10) |
| $SEG_L$ $GT_H$ | 0.86(0.05) | 0.91(0.03) | 0.90(0.05) | 0.76(0.08) | 0.96(0.20) | 0.89(0.04) | 0.88(0.05) | 0.88(0.03) |

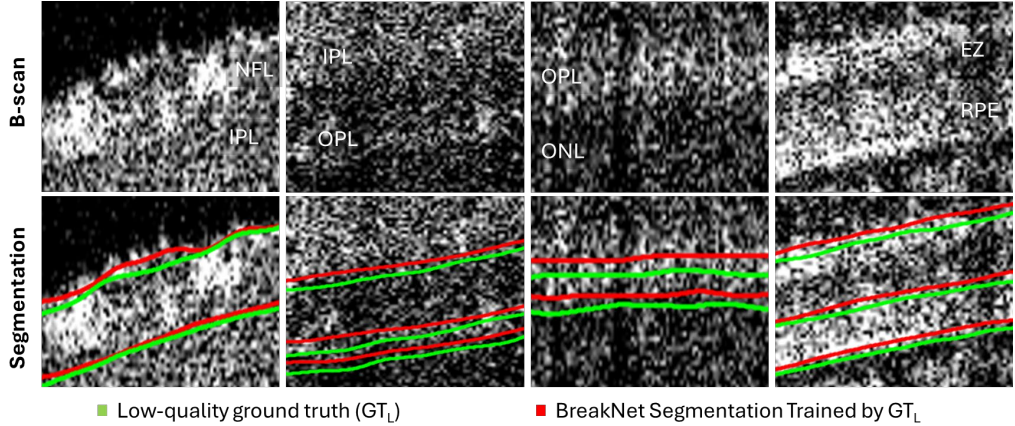

**Fig. 4.** Representative vis-OCT B-scan images of rat retinas in zoom-in view overlaid with layer segmentations (green: low-quality ground truth, $GT_L$, red: BreakNet segmentation trained by the low-quality ground truth).

*3.5 Evaluation of BreakNet on Mouse Retinas*

To further assess the performance of our proposed method, we re-trained BreakNet using vis-OCT images from mouse retinas. Since the mouse retinas have smaller eyes and caliber of retinal major vessels, their blood vessel shadows, and layer discontinuity are less pronounced than those in rats (Fig. 5). As a result, the Dice and IoU values were improved to 0.91±0.01 and 0.84±0.02, respectively in mice, indicating the effectiveness of BreakNet in segmenting retinal layers across multiple species.

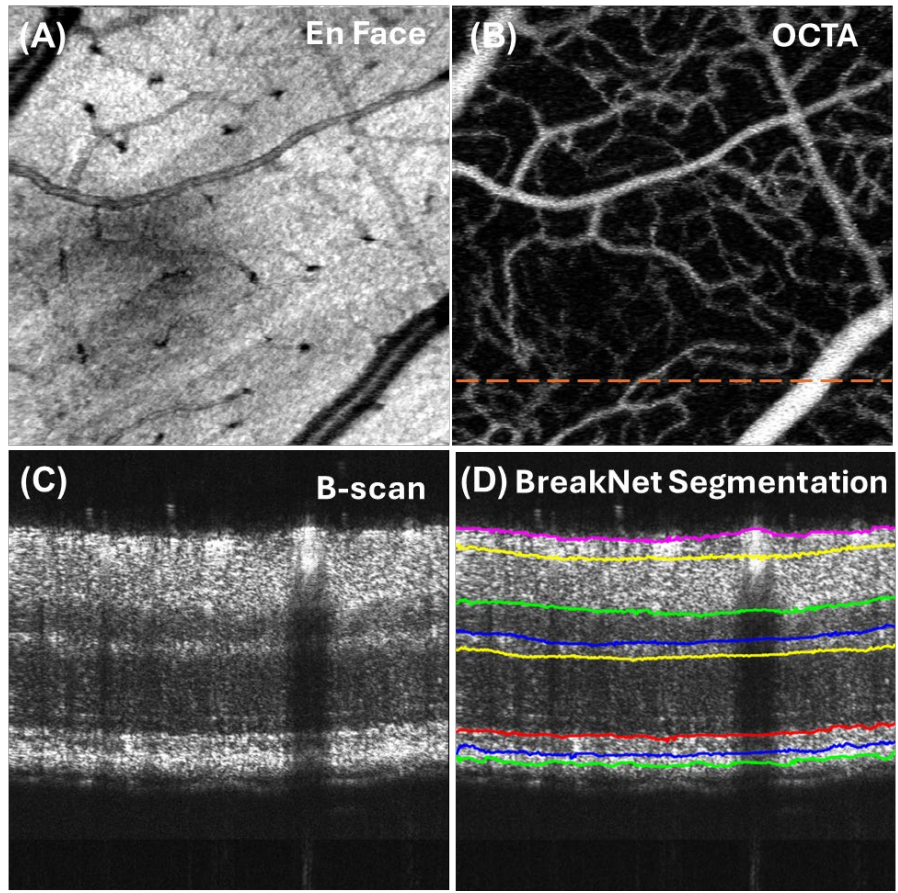

**Fig. 5.** Representative vis-OCT images of mice retina showing A) *en face* image of OCT, B) *en face* image of OCTA, C) cross-sectional B-scan image of OCT, and D) B-scans overlaid with segmentation by BreakNet.

### 4. Discussion

Vis-OCT has shown promise in achieving high-resolution and functional retinal imaging, surpassing the capabilities of standard near-infrared OCT. However, the inherent challenge of blood vessel shadow artifacts, due to strong hemoglobin absorption, complicates accurate retinal layer segmentation. These artifacts create significant discontinuities in the OCT signal beneath blood vessels, obstructing layer delineation. In this study, we propose a novel deep-learning architecture, BreakNet, designed to overcome these challenges. Aiming to improve segmentation accuracy in the presence of blood vessel shadows, BreakNet integrates multi-scale convolutional and Transformer-based blocks to extract both local and global features. In regions where edge information fades, or where noise and other artifacts disrupt the edges and textures, the global feature extraction capability compensates for the weakened locality. This global understanding enables the model to learn high-level features, resulting in accurate segmentation even in areas where local information is reduced or discontinued by vessel shadow artifacts.

We evaluated BreakNet on vis-OCT images of rodent retinas, and our model demonstrated superior segmentation performance compared to state-of-the-art methods. We validated its performance through ablation studies, testing its usefulness with low-quality ground truth, and assessing its generalization across species. BreakNet's superior performance in addressing the challenges posed by blood vessel shadows in vis-OCT images has significant implications for retinal imaging and analysis. By enabling accurate layer segmentation despite artifacts, BreakNet facilitates better quantification and analysis of retinal structures, potentially improving the diagnosis and monitoring of retinal diseases.

In medical image processing, incorporating expert knowledge into deep-learning models can significantly enhance segmentation performance [26]. When generating ground truth, expert graders tend to delineate layers by maintaining consistent thickness and curvature in other regions or B-scans. In this study, vision transformers successfully utilized this knowledge. As shown in Fig. 3, the importance of these features is particularly evident in challenging images, where a considerable amount of information is obscured or weakened by large vessel shadows. Our results demonstrate that BreakNet can effectively learn and mimic this expert ability for accurate segmentation. As mentioned earlier, the layer discontinuity in NIR-OCT human retinal images is not as strong as that observed in rats with vis-OCT, even though humans have larger retinal vessels than rodents. Therefore, we anticipate that BreakNet can effectively segment NIR-OCT human retinal images. To validate this hypothesis, we trained BreakNet from scratch using the OCTA500 dataset [27], which includes normal and pathological retinas acquired with NIR-OCT at 840 nm. As expected, BreakNet confidently segmented retinal layers in normal healthy subjects, without confusing the dark regions caused by vessel shadows with cystoid fluid (Fig. 6). More importantly, BreakNet successfully identified the boundaries in challenging cases, such as vitreous shadow, hard exudates, and retinoblastoma, where significant layer discontinuities were present (Fig. 6).

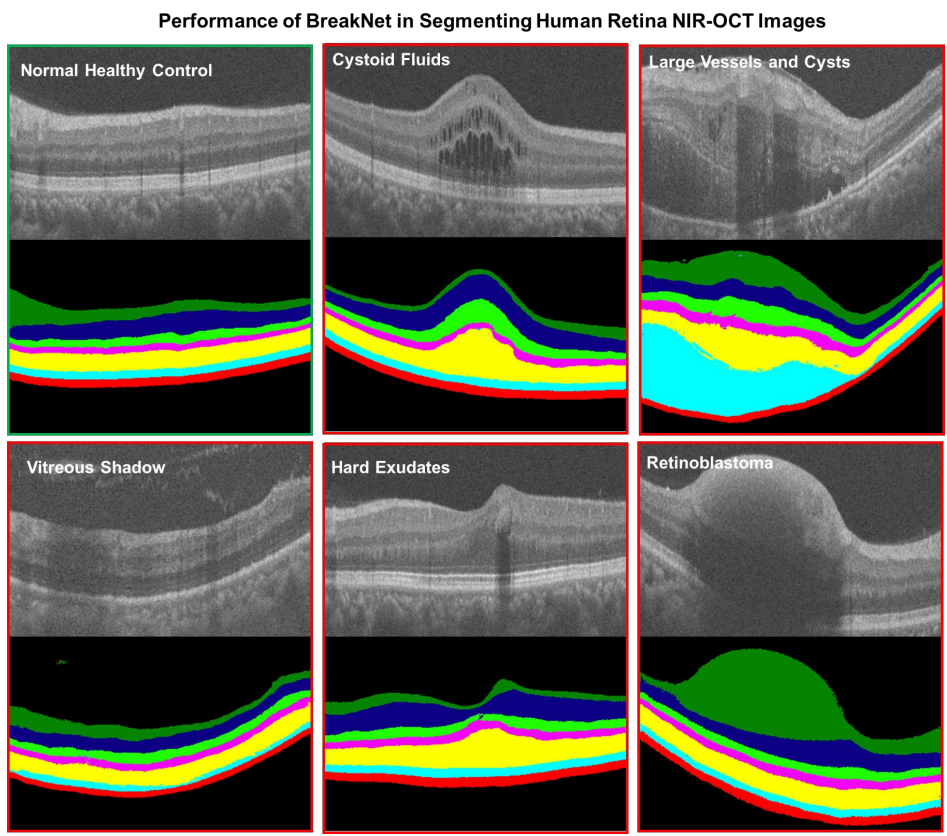

**Fig. 6.** Representative NIR-OCT images of human retinas showing raw B-scans and segmentation results by BreakNet. Images include a normal case and various pathological conditions: cystoid fluid damage, large vessel shadows along with cysts, vitreous shadow, hard exudates, and retinoblastoma. Normal cases are highlighted with green bounding boxes, and challenging cases are marked with red bounding boxes.

Future research could explore several avenues to further enhance BreakNet's capabilities. As vis-OCT can visualize more than 8 layers in the retinas [7, 28], it would be beneficial to improve the model's performance in segmenting sub-laminar structures by incorporating advanced networks and/or a sequence of B-scans. Another direction could involve refining the architecture to further reduce computational complexity while maintaining segmentation accuracy. Additionally, extending BreakNet to other imaging modalities and anatomical regions could demonstrate its versatility and broaden its application scope. Integrating BreakNet with advanced techniques such as transfer learning and unsupervised learning could leverage pre-trained models on large-scale datasets to enhance performance in scenarios with limited annotated data. Exploring real-time segmentation capabilities could also pave the way for clinical applications, where rapid and accurate analysis is crucial.

## 5. Conclusion

BreakNet represents a significant advancement in retinal layer segmentation for vis-OCT images, particularly in overcoming the challenges posed by blood vessel shadows. Its robust performance demonstrated through comprehensive evaluations and comparative analyses, underscores its potential for improving retinal imaging and analysis. The ability to generalize across species and maintain accuracy with limited-quality ground truths further enhances its practical applicability. Future developments and extensions of BreakNet hold promise for broadening its impact in medical imaging and beyond.

## Acknowledgments

We appreciate the funding from the Knight Templar Eye Foundation (SP), Alcon Research Institute (SP), and Eye & Ear Foundation of Pittsburgh (SP and BW). We also acknowledge support from NIH CORE Grant P30 EY08098 and an unrestricted grant from Research to Prevent Blindness to the Department of Ophthalmology.

## Disclosures

The authors declare no conflicts of interest.

## Data availability

Data underlying the results presented in this paper are not publicly available at this time but may be obtained from the authors upon reasonable request.

## References

1. Yi, J., et al., *Visible-light optical coherence tomography for retinal oximetry.* Optics Letters, 2013. **38**(11): p. 1796-1798.
2. Yi, J., et al., *Visible light optical coherence tomography measures retinal oxygen metabolic response to systemic oxygenation.* Light: Science & Applications, 2015. **4**(9): p. e334-e334.
3. Pi, S., et al., *Retinal capillary oximetry with visible light optical coherence tomography.* Proceedings of the National Academy of Sciences, 2020. **117**(21): p. 11658-11666.
4. Wang, J., et al., *A dual-channel visible light optical coherence tomography system enables wide-field, full-range, and shot-noise limited human retinal imaging.* Communications Engineering, 2024. **3**(1): p. 21.
5. Miller, D.A., et al., *Visible-Light Optical Coherence Tomography Fibergraphy of the Tree Shrew Retinal Ganglion Cell Axon Bundles.* IEEE Transactions on Medical Imaging, 2024: p. 1-1.
6. Cai, Z., et al., *Multiscale imaging of corneal endothelium damage and Rho-kinase inhibitor application in mouse models of acute ocular hypertension.* Biomedical Optics Express, 2024. **15**(2): p. 1102-1114.
7. Zhang, T., et al., *Visible light OCT improves imaging through a highly scattering retinal pigment epithelial wall.* Optics Letters, 2020. **45**(21): p. 5945-5948.
8. Chauhan, P., A.M. Kho, and V.J. Srinivasan, *From Soma to Synapse: Imaging Age-Related Rod Photoreceptor Changes in the Mouse with Visible Light OCT.* Ophthalmology Science, 2023. **3**(4): p. 100321.
9. Soetikno, B.T., et al., *Visible-light optical coherence tomography oximetry based on circumpapillary scan and graph-search segmentation.* Biomedical Optics Express, 2018. **9**(8): p. 3640-3652.
10. Gopal, B.D., et al., *Automatic retinal layer segmentation of visible-light optical coherence tomography images using deep learning.* Investigative Ophthalmology & Visual Science, 2022. **63**(7): p. 2069 – F0058-2069 – F0058.
11. Guo, Y., et al., *An end-to-end network for segmenting the vasculature of three retinal capillary plexuses from OCT angiographic volumes.* Biomedical Optics Express, 2021. **12**(8): p. 4889-4900.

12. Ye, T., J. Wang, and J. Yi, *Deep learning network for parallel self-denoising and segmentation in visible light optical coherence tomography of the human retina.* Biomedical Optics Express, 2023. **14**(11): p. 6088-6099.
13. Vaswani, A., et al., *Attention is all you need.* Advances in neural information processing systems, 2017. **30**.
14. Chen, J., et al., *Transunet: Transformers make strong encoders for medical image segmentation.* arXiv preprint arXiv:2102.04306, 2021.
15. Liu, Z., et al. *Swin transformer: Hierarchical vision transformer using shifted windows*. in *Proceedings of the IEEE/CVF international conference on computer vision*. 2021.
16. Tan, Y., et al., *Retinal layer segmentation in OCT images with boundary regression and feature polarization.* IEEE Transactions on Medical Imaging, 2023.
17. Cao, G., et al., *A single-step regression method based on transformer for retinal layer segmentation.* Physics in Medicine & Biology, 2022. **67**(14): p. 145008.
18. Oktay, O., et al., *Attention u-net: Learning where to look for the pancreas. arXiv 2018.* arXiv preprint arXiv:1804.03999, 1804.
19. Cao, G., et al., *Self-attention CNN for retinal layer segmentation in OCT.* Biomedical Optics Express, 2024. **15**(3): p. 1605-1617.
20. Lee, Y., et al. *Mpvit: Multi-path vision transformer for dense prediction*. in *Proceedings of the IEEE/CVF conference on computer vision and pattern recognition*. 2022.
21. Yu, W., et al. *Metaformer is actually what you need for vision*. in *Proceedings of the IEEE/CVF conference on computer vision and pattern recognition*. 2022.
22. Xu, W., et al. *Co-scale conv-attentional image transformers*. in *Proceedings of the IEEE/CVF international conference on computer vision*. 2021.
23. Chiu, S.J., et al., *Kernel regression based segmentation of optical coherence tomography images with diabetic macular edema.* Biomedical optics express, 2015. **6**(4): p. 1172-1194.
24. Karri, S., D. Chakraborthi, and J. Chatterjee, *Learning layer-specific edges for segmenting retinal layers with large deformations.* Biomedical optics express, 2016. **7**(7): p. 2888-2901.
25. Ronneberger, O., P. Fischer, and T. Brox. *U-net: Convolutional networks for biomedical image segmentation*. in *Medical image computing and computer-assisted intervention–MICCAI 2015: 18th international conference, Munich, Germany, October 5-9, 2015, proceedings, part III 18*. 2015. Springer.
26. Xie, X., et al., *A survey on incorporating domain knowledge into deep learning for medical image analysis.* Medical Image Analysis, 2021. **69**: p. 101985.
27. Li, M., et al., *OCTA-500: A Retinal Dataset for Optical Coherence Tomography Angiography Study.* arXiv e-prints, 2020: p. arXiv: 2012.07261.
28. Wang, L., J.A. Sahel, and S. Pi, *Sub2Full: split spectrum to boost optical coherence tomography despeckling without clean data.* Optics Letters, 2024. **49**(11): p. 3062-3065.